\def \SAIT #1 #2 {{\em Mem.\ Soc.\ Astron.\ It.\/} {\bf #1}, #2}
\def \MESS #1 #2 {{\em The Messenger\/} {\bf #1}, #2}
\def \ASTRNACH #1 #2 {{\em Astron. Nach.\/} {\bf #1}, #2}
\def \AAP #1 #2 {{\em Astron. Astrophys.\/} {\bf #1}, #2}
\def \AAL #1 #2 {{\em Astron. Astrophys. Lett.\/} {\bf #1}, L#2}
\def \AAR #1 #2 {{\em Astron. Astrophys. Rev.\/} {\bf #1}, #2}
\def \AAS #1 #2 {{\em Astron. Astrophys. Suppl. Ser.\/} {\bf #1}, #2}
\def \AJ #1 #2 {{\em Astron. J.\/} {\bf #1}, #2}
\def \ANNREV #1 #2 {{\em Ann. Rev. Astron. Astrophys.\/} {\bf #1}, #2}
\def \APJ #1 #2 {{\em Astrophys. J.\/} {\bf #1}, #2}
\def \APJL #1 #2 {{\em Astrophys.. J. Lett.\/} {\bf #1}, L#2}
\def \APJS #1 #2 {{\em Astrophys. J. Suppl.\/} {\bf #1}, #2}
\def \APSS #1 #2 {{\em Astrophys. Space Sci.\/} {\bf #1}, #2}
\def \ASR #1 #2 {{\em Adv. Space Res.\/} {\bf #1}, #2}
\def \BAIC #1 #2 {{\em Bull. Astron. Inst. Czechosl.\/} {\bf #1}, #2}
\def \JSQRT #1 #2 {{\em J. Quant. Spectrosc. Radiat. Transfer\/} {\bf #1}, #2}
\def \MN #1 #2 {{\em Mon. Not. R. Astr. Soc.\/} {\bf #1}, #2}
\def \MEM #1 #2 {{\em Mem. R. Astr. Soc.\/} {\bf #1}, #2}
\def \PLR #1 #2 {{\em Phys. Lett. Rev.\/} {\bf #1}, #2}
\def \PASJ #1 #2 {{\em Publ. Astron. Soc. Japan\/} {\bf #1}, #2}
\def \PASP #1 #2 {{\em Publ. Astr. Soc. Pacific\/} {\bf #1}, #2}
\def \NAT #1 #2 {{\em Nature\/} {\bf #1}, #2}
\def\msol{{M}_{\odot}}

\documentstyle{memsait}
\input psfig.tex
\begin{opening}
\title{CLUSTERS OF GALAXIES: MULTIFREQUENCY OBSERVATIONS VERSUS THEORIES}
\author{SABINE SCHINDLER}
\institute{Astrophysics Research Institute, 
               Liverpool John Moores University, 
	       Twelve Quays House,
               Birkenhead L41 1LD,
               United Kingdom\\
}
\date{} 
\end{opening}

\begin{document}

\oddpagefooter{}{}{} 
\evenpagefooter{}{}{} 
\ 
\bigskip

\begin{abstract}
Clusters of galaxies are large gravitationally bound systems which
consist of several observable components: hundreds of galaxies, hot
gas between the galaxies and sometimes relativistic particles. These
components are emitting in different wavelengths from radio to
X-rays. We show that the combination of 
observations at different frequencies and also theoretical models 
is  giving now a
comprehensive picture of these massive objects. Topics presented here
include  cluster masses, 
baryon fractions, the dynamical state of clusters, the physical
processes in clusters and 
cosmological parameters derived from cluster observations. 
\end{abstract}

\section{Introduction}

Clusters of galaxies were first detected as large concentrations of
galaxies. With the advent of X-ray astronomy also X-ray
emission from galaxy clusters was found. This emission could be
explained as thermal bremsstrahlung from hot gas filling the whole
potential well of the cluster (e.g. Sarazin 1986). 
Moreover, in many clusters radio
emission was found. This radio emission is synchrotron emission from
relativistic particles. Table~1 summarises the main cluster
components together with their mass fractions. While galaxies
contribute only $3-5\%$ to the cluster mass, the contribution of the
gas is considerably more ($10-30$\%; see e.g. Arnaud \& Evrard 1999; 
Ettori \& Fabian 1999). But, as these two contributions
are by far not 100\%, it is concluded that most of the mass is in
form of dark matter, i.e. not 
directly observable. Therefore mass determinations, which are indirect
measurements of the dark matter, are so particularly interesting in
clusters. 

The list of components summarises only the main cluster components and
main frequencies, at which clusters are observed. 
Of course this list is not complete, and clusters are also observable
at other frequencies, e.g. non-thermal,
hard X-ray emission was detected in some clusters 
(see Fusco-Femiano et al. 1999).
In the following several fields of cluster research are
presented. Also this compilation cannot be complete with only 
limited space available. Only a few
topics could be selected and handled very briefly. 
Preference was given to currently very
active fields which combine observations in two or more wavelengths.

\vspace{1cm} 
\centerline{\bf Tab. 1 - Main cluster components}

\begin{table}[h]
\hspace{1.5cm} 
\begin{tabular}{|l|c|c|}
\hline
                      & mass fraction & observable in \\
\hline
galaxies              & $3-5\%$       & optical       \\
intra-cluster gas     & $10-30\%$     & X-rays (thermal bremsstrahlung)\\
relativistic particles& -             & radio  (synchrotron emission)\\
dark matter           & rest          & -             \\
\hline
\end{tabular}
\end{table}

\section{Optical observations}

Historically, optical observations were the first cluster
observations. Already from the first cluster catalogues (e.g. Abell
1958) a rough estimate of the richness and the morphology of cluster
could be obtained. With the additional information of the velocity
of the galaxies three aspects can be addressed: (1) the
redshift measurement yields the three-dimensional distribution of
clusters, which is very important for cosmological studies, i.e. the
determination of distribution functions. (2) The distribution of
velocities within a cluster gives information about the internal
dynamics, i.e. the collision of two subclusters can show up as
a broadened velocity distribution 
(e.g. Binggeli et al. 1993). (3) With the assumption of virial
equilibrium the velocity dispersion (= the standard deviation of the
distribution) yields a measure for the total cluster mass (see
e.g. Carlberg et al. 1996; Girardi et al. 1998). In this way  
Zwicky found already in 1933 that not all the cluster mass can be
contained in the galaxies. Detailed spectroscopic and morphological
studies of distant cluster galaxies provides interesting information on
cluster formation and evolution (e.g. Stanford et al. 1998, van Dokkum
et al. 1998).
The morphological types of galaxies
are not distributed in uniformly, but the higher the galaxy density
the larger is the fraction of ellipticals (Dressler 1980). This holds
not only 
for a comparison of the field galaxies with cluster galaxies, but also
for cluster galaxies at different distances to the cluster
centre. This effect can be well seen e.g. in the Virgo cluster
(Binggeli et al. 1987; Schindler et al. 1999; see Fig.~1). The
explanation for this density-morphology relation is
interaction between galaxies and interaction of galaxies with the
intra-cluster gas. A beautiful example of the latter can be seen again
in the Virgo cluster in HI: two galaxies are stripped off their cool
gas when approaching the cluster centre (Cayatte et al. 1990).

\begin{figure}
\begin{tabular}{cc}
\psfig{figure=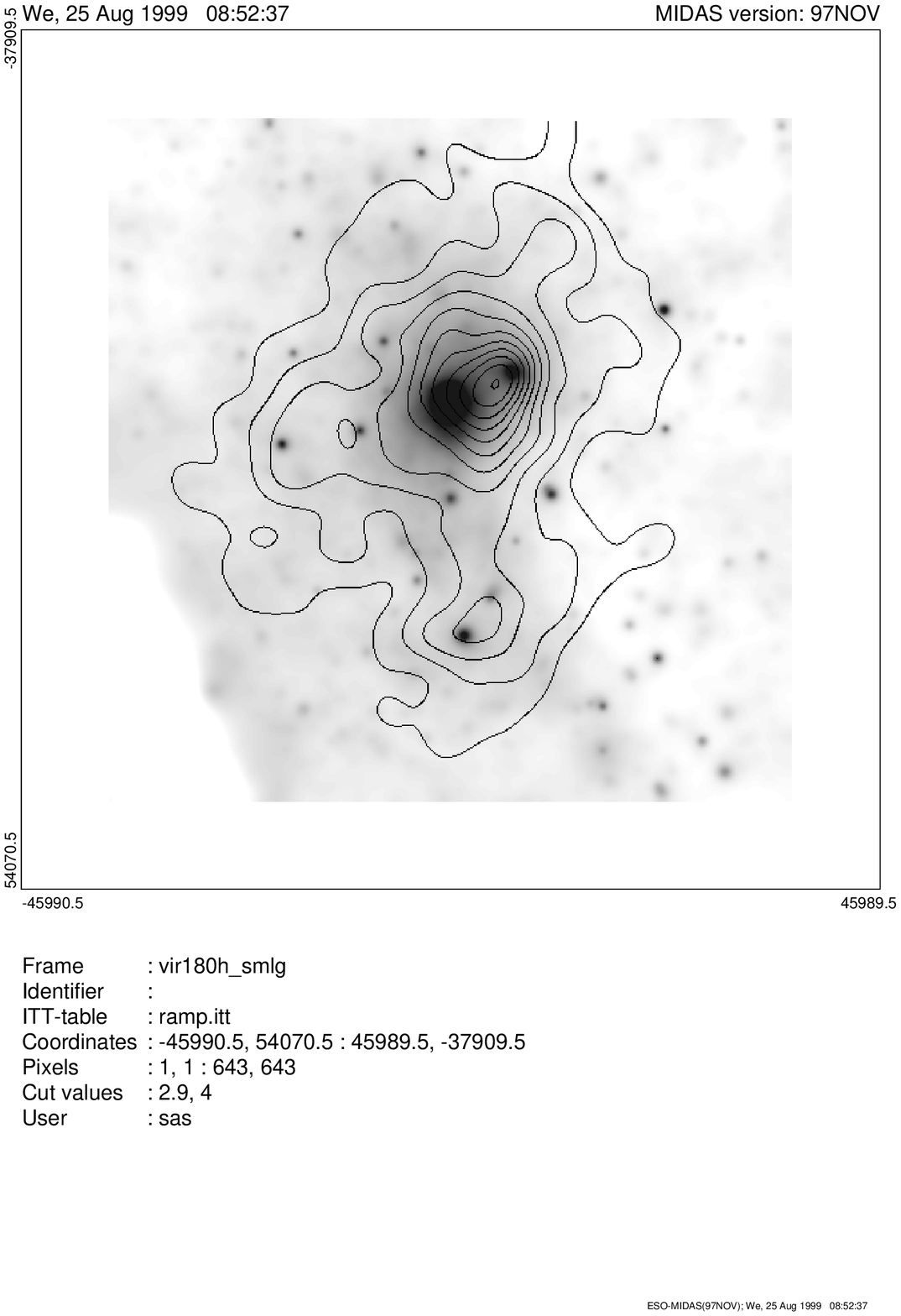,width=6.2cm,clip=} &
\psfig{figure=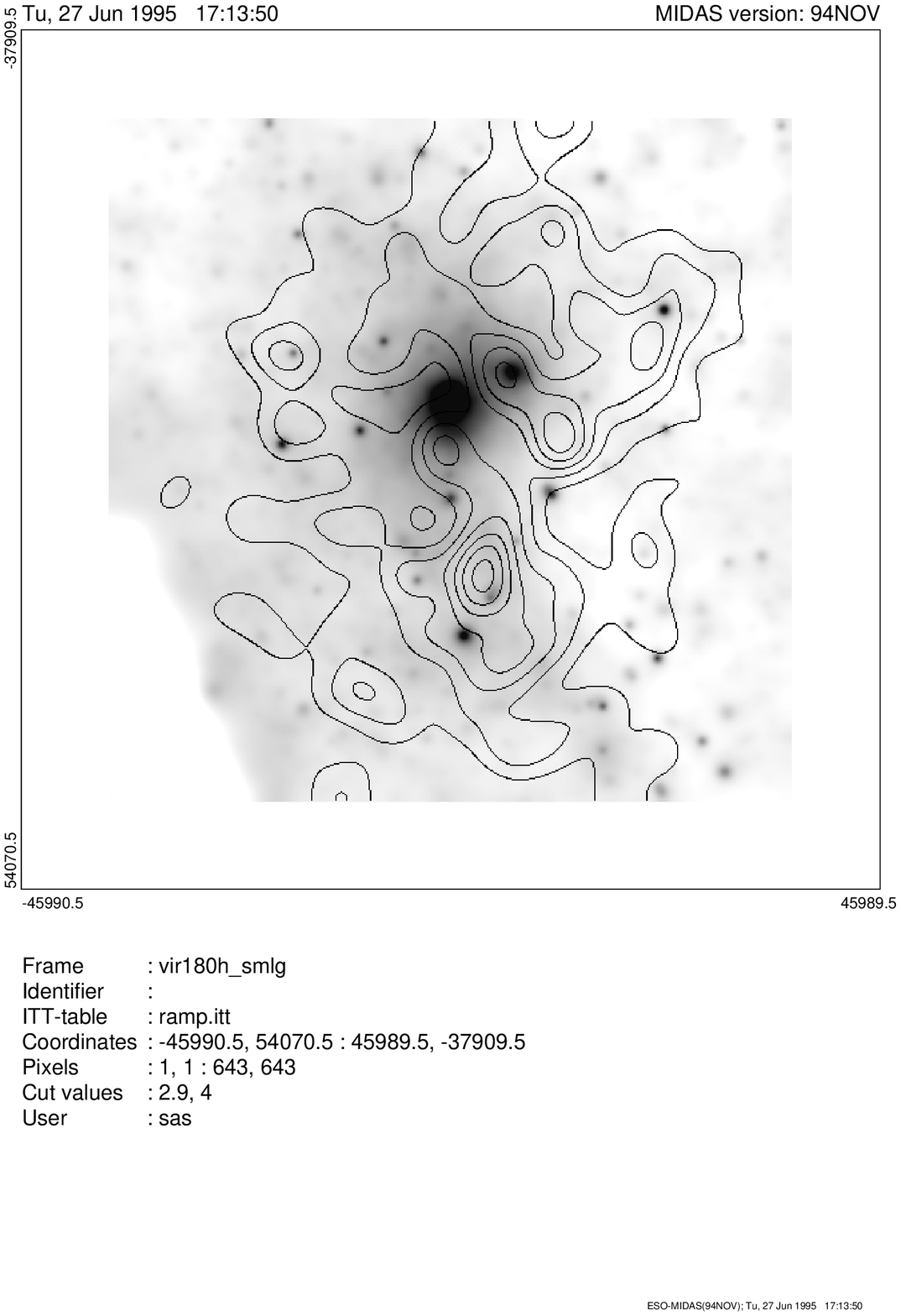,width=6.2cm,clip=} \\
\end{tabular}
\caption[h]{X-ray emission of the intra-cluster gas (greyscale image)
from the Virgo cluster in the ROSAT
All-Sky Survey. Superposed on the X-ray image are contours of constant
galaxy number
density of different morphological types. The dwarf elliptical
galaxies (left) show a very centrally concentrated distribution -- 
similar to the distribution of the gas.
The spirals and irregulars (right) are much more dispersed and the
density maxima do not coincide with the X-ray maxima.
}
\end{figure}

A currently very active field is gravitational lensing. Since the
first arcs in clusters were discovered more than 10 years ago (Lynds \&
Petrosian 1986; Soucail 1987) gravitational lensing 
was used as a way to determine cluster masses. Two
methods to determine the mass can be distinguished: strong lensing and
weak lensing. Strong lensing uses the giant arcs which are distorted
images of background galaxies, see e.g. the
beautiful HST images of Cl0024+1654 (Colley et al. 1996) and A2218
(Kneib et al. 1996). With this method only the mass contained in a
volume within the arc radius can be measured, i.e. it is restricted to
the central part of the cluster. 
Weak lensing on the other hand uses the systematic
elongation of all background galaxies. With a few mathematical
operations -- a method pioneered by Kaiser \& Squires (1993) -- this can
be directly transformed into a mass distribution without treating 
different subclusters or single galaxies separately. The difficulties for
this method are the distinction of background galaxies and the
normalisation of the mass, because the mass
distribution cannot be measured out to border of the cluster
due to the limited CCD sizes. An example for a
weak lensing analysis for the cluster A2218 and a comparison of the
different mass determination methods can be seen in Squires et al. (1996). 
For a recent reviews on lensing in clusters see Hattori et al. (1999) and for
lensing in general see Wambsganss (1998).

\section{X-ray observations}

The gas between the galaxies is so hot, that it is emitting thermal
bremsstrahlung in X-rays. This hot gas is filling the whole cluster
potential and  is therefore a good tracer for deep potential wells. 
As the thermal bremsstrahlung is proportional to the square of the gas
density, X-ray selected clusters are much less affected by projection
effects than optically selected clusters. Furthermore, the morphology
of a cluster can be seen much better in X-rays, even for distant
clusters. 
Two examples of clusters with very different morphologies (see Fig.~2)
at similar redshifts ($z\approx 0.4$)
are Cl0939+4713 -- a cluster with two subclumps each of
them showing even some internal structure (Schindler et al. 1998) --
and RXJ1347-1145 -- the most X-ray luminous cluster found so far with
a very centrally concentrated X-ray emission (Schindler et al. 1997).
The morphologies are important for cosmology.
In a low $\Omega$ universe the merging processes should stop earlier
so that more clusters with virialised X-ray emission are expected.
Therefore the fraction of virialised clusters at a certain redshift 
can be used to constrain the mean density of the universe $\Omega$ 
(for theory see Richstone et al. 1992; for an application to
observations see Mohr et al. 1995). 

\begin{figure}
\begin{tabular}{cc}
\psfig{figure=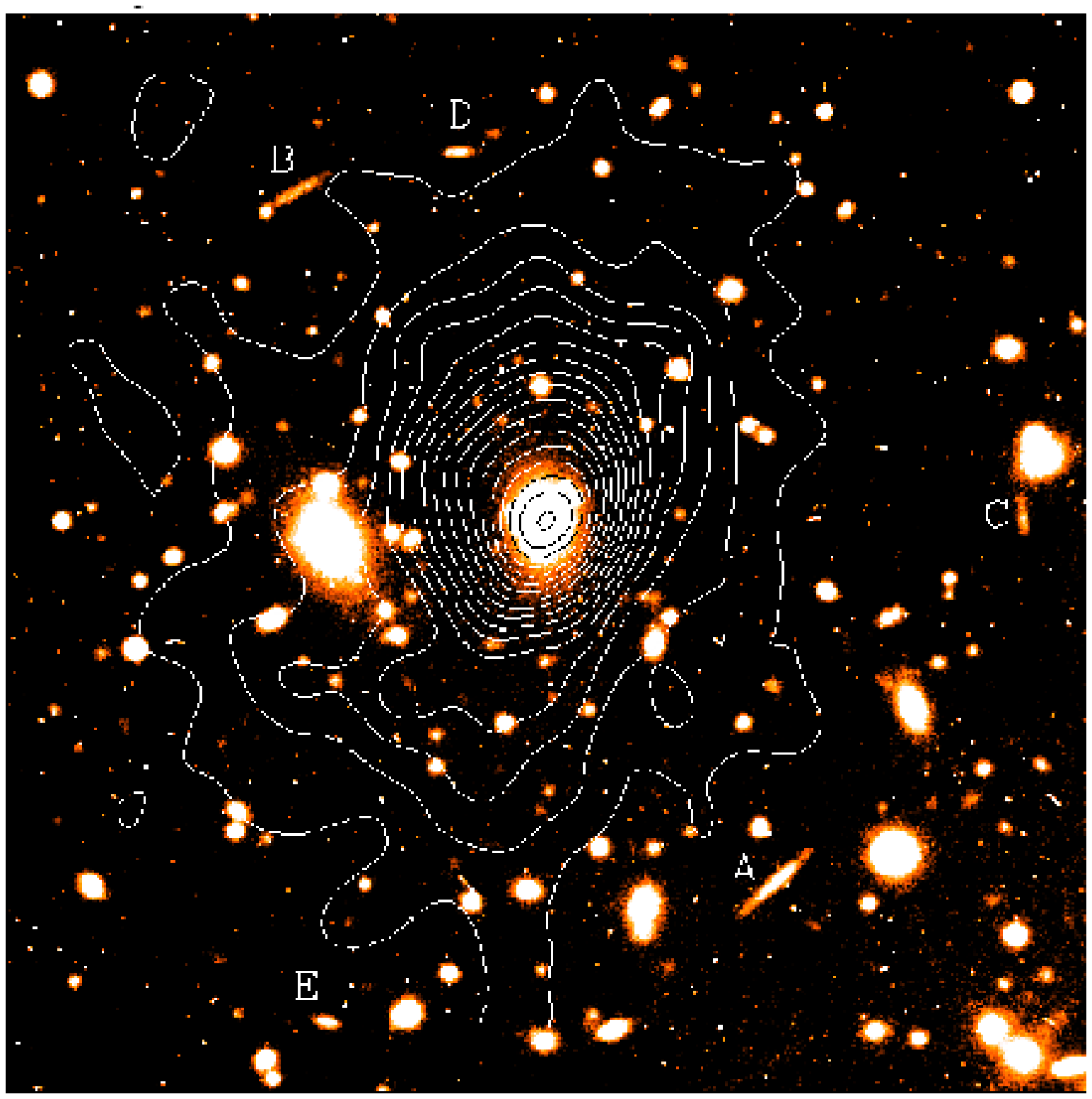,width=6.3cm,clip=} &
\psfig{figure=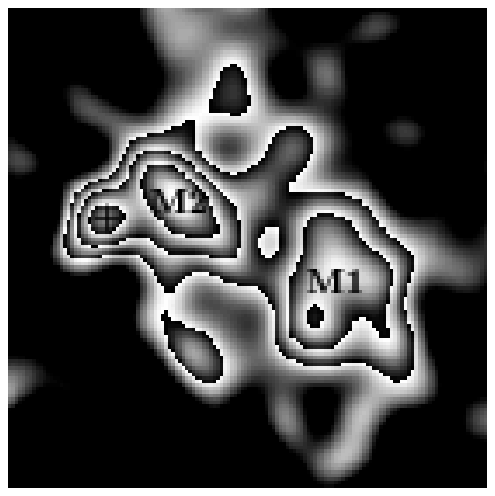,width=6.1cm,clip=} \\
\end{tabular}
\caption[h]{Two clusters at redshift $z\approx 0.4$ with very different
X-ray morphology. RXJ1347-1145 (left) -- the most X-ray luminous
cluster -- shows a very compact gas
distribution in X-rays (contours). The contours are superposed on an 
optical image. Several arcs around the centre (marked by letters) show
that the cluster is acting as a gravitational lens.
The cluster Cl0939+4713 (right) shows a very
different gas distribution. Two subclusters marked with M1 and M2 are
visible, which have even some internal structure. The maximum marked
with a ``plus'' sign is not cluster emission, but is caused by a
background quasar at $z=2$. The size of the two images is very
different: the image of RXJ1347-1145 is $1.5\times1.5$ arcminutes$^2$,
while image of Cl0939+4713 is $3.7\times3.7$ arcminutes$^2$.
}
\end{figure}

While recent morphological studies were carried out mainly with ROSAT
observations, the Japanese satellite ASCA
was much used because of its spectral capabilities. An important
parameter determined by X-ray spectra is the gas temperature.
The temperature is typically between 1
and 10 keV being in good agreement with the depth of the potential
well. Temperature maps, which are still very coarse due to the limited
spatial resolution of ASCA, show that many clusters are not isothermal
(Markevitch 1998). The temperature distribution is another way to
determine the dynamical state of a cluster. As it was shown by
hydrodynamic simulations (Schindler \& M\"uller 1993) the
temperature structure shows very clearly the different stages of a
merger, e.g. the hot, compressed gas between two subclusters shortly
before they collide or the shock waves emerging after the collision
as steep temperature gradients.

As not all the
ions are completely ionised at these temperatures line emission can be
observed. In most of the clusters Fe lines are visible, but sometimes
also Si and other elements are detectable.
The metallicities (calculated from the Fe lines) 
are typically in the range
between 0.2 and 0.5 in solar units (e.g. Mushotzky \&
Loewenstein 1997; Tsuru et al. 1997; Fukazawa et al. 1998). 
Obviously, the gas cannot be
purely of primordial origin but must have been enriched 
by nucleosynthesis processes in cluster galaxies. No
(strong) evolution of the metallicities with time has been found out
to a redshift of 1 (Schindler 1999). This result is in agreement with
optical observations of cluster galaxy evolution (e.g. Stanford et
al. 1998)
and theoretical models (Martinelli et al. 1999), which both find no
metal enrichment of the intra-cluster medium for redshifts smaller
than 1, i.e. the enrichment must take place relatively early. 
A famous example for a high-redshift and high-metallicity cluster 
is the cluster AXJ2019+112 at a
redshift of $z=1$, which has a puzzling iron abundance of more than the
solar value (Hattori et al. 1997).

X-ray observations provide another method the measure the cluster mass.
With the assumption of hydrostatic equilibrium the mass
can be estimated  simply from the temperature profile and the gas
density profile (for applications of this method on a large sample of
clusters see e.g. Arnaud \& Evrard 1999; Ettori \& Fabian 1999). 
Numerical simulations showed that hydrostatic
equilibrium is a good assumption to get reasonable mass estimates in
roughly relaxed clusters
(Evrard et al. 1996; Schindler 1996). 
Masses for typical clusters are
of the order of $10^{15}\msol$ when measured out to the virial radius. A
comparison of the masses determined by the different methods shows
that the virial mass and the X-ray mass are in general in agreement, but the
lensing mass (in particular the mass from strong lensing) is sometimes
up to a factor of three higher (see e.g. the comparison in Squires et
al. 1996 or Schindler et al. 1997). In these comparisons it is
important to take into account that the mass is measured in different
volumes: in X-rays the mass is measured in a spherical volume,
while the gravitational lensing effect is sensitive to all the mass along
the line-of-sight, i.e. it measures the mass in a cylindrical volume
around the cluster.
Therefore, the measurements can give different results, although they
are all correct.
As the measurements for all the methods mentioned here are getting
better and the problems of the individual methods (like
e.g. projection effects) are better understood, it is probable that
all the methods are going to converge in the end.

From the X-ray observations also the mass of the intra-cluster gas 
can be
determined. Together with the mass in the galaxies this yields an 
average baryon fraction of about 20\%. For an $\Omega=1$ universe this
value is at least 3 times larger than allowed by primordial
nucleosynthesis -- a famous discrepancy termed   
``baryon catastrophe'' (White et al. 1993). The easiest way out of the
problem seems now a cosmological model with a low $\Omega$.

A comparison of the gas distribution and the dark matter distribution
shows that the gas distribution is generally more extended (e.g. David et
al. 1995). 
Obviously, cluster evolution is not completely a self-similar process,
but physical processes taking place in the gas must be taken into
account, like e.g. energy input by supernovae, galactic winds or
ram-pressure stripping (see e.g. Metzler \& Evrard 1997; Cavaliere et
al. 1998a). As the gas distribution is relatively more extended in
less massive clusters (see Schindler 1999) these heating processes
must be more efficient in less massive clusters.

Clusters as the largest bound objects in the universe are very good
tracers for large-scale structure. 
They can be used for various cosmological tests. Distribution
functions like the luminosity function (e.g. De Grandi et al. 1999) or
the correlation function (e.g. Guzzo et al. 1999) preferably 
of an X-ray selected
cluster sample (i.e. a mass selected sample) 
can be used to constrain cosmological parameters. 
Also correlations between X-ray quantities (e.g. between the X-ray
luminosity, the temperature and the cluster mass) can be used to test different
cosmological models because their relations as well as the evolution
of these relations depend on cosmological
parameters (Oukbir \& Blanchard 1992; Bower 1997;
Cavaliere et al. 1998b, Eke et al. 1998; Schindler 1999).

\section{Radio observations}

Radio emission has been found in many galaxies clusters. Two different
kinds of radio emission can be distinguished: diffuse emission and
emission associated with galaxies. The latter one (see Owen \& Ledlow
1997 for many examples) can be used to determine the relative
motion of the intra-cluster gas and head-tail galaxies 
by their radio morphology (O'Donoghue et al. 1993;
Sijbring \& de Bruyn 1998). Furthermore, the pressure of the
intra-cluster gas can be estimated from the radio lobe
expansion into the gas (e.g. Eilek et al. 1984; Feretti et al. 1990). 
Observations of the
rotation measure of sources in or behind a cluster provide the
possibility to determine the cluster magnetic field. Typically values 
between 0.1 $\mu$G up to few $\mu$G are found (e.g. Feretti et al. 1995).

In several clusters diffuse radio emission could be detected
(e.g. Giovannini et al. 1999). If the diffuse emission is 
located in the central parts and has a roughly spherical shape, 
it is  called radio halo, see e.g. the Coma cluster
(Giovannini et al. 1993).
In other clusters the radio emission is
situated in the outer parts and has usually elongated shapes. These
sources are called relics, see e.g. A3667 (R\"ottgering et al. 1997).
Although it was previously assumed that the central and non-central 
sources are different kinds of sources, it is now probable that they 
have the same origin. A possible
explanation for the emission could be merging of subclusters.
In such mergers turbulence
and shocks are produced which can provide the necessary energy to
reaccelerate particles and to amplify the magnetic field. The
discovered correlations  between the halo size, the radio power, the X-ray
luminosity  and the gas temperature of the host cluster
support this theory: The collision of more massive 
clusters (= higher X-ray luminosity and higher gas temperature) would
provide more energy for the radio halo.

Finally, a very exciting field, which uses a combination of radio and
X-ray observations, is the distance determination by the
Sunyaev-Zel'dovich effect (Sunyaev \& Zel'dovich 1972). 
When photons of the cosmic microwave background pass through the
hot gas of a cluster, they are scattered to slightly higher
energies, i.e. inverse Compton scattering. That means the blackbody
spectrum of the CMB appears slightly shifted when observed in
direction of a cluster. This results in an increment or a decrement
depending on what side of the blackbody spectrum the observations are
done. This in- or decrement is proportional to the intra-cluster gas
density, while the X-ray emission is proportional to the square of the
density. These two different dependences allow to estimate the
physical size of the cluster, while the angular size of
the cluster can be measured easily. The combination of physical and
angular size provides a direct measurement of the distance of the
cluster. The problem of this method is 
that the physical size is measured along
the line-of-sight, while the angular size is measured perpendicular to
the line-of-sight, i.e. if the cluster is elongated the derived
distance is wrong. To avoid this problem currently whole samples of
clusters are measured (e.g. Carlstrom et al. 1999), 
because for many clusters this effect is
expected to average out. Furthermore, the Sunyaev-Zel'dovich effect
can also be used to study the distribution of the intra-cluster gas.
For a recent review on the Sunyaev-Zel'dovich effect see Birkinshaw
(1999). 

\section{Conclusions}

The study of clusters of galaxies became a very active field in recent
years through the development of new techniques (e.g. gravitational
lensing) and powerful instruments (e.g. ROSAT, HST). 
Observations in all wavelengths and the comparison with theory
taught us a lot about cluster components,
cluster dynamics and the
physics in clusters. A particularly interesting aspect has opened up in the
last years with the use of clusters
as probes for cosmology. The new instruments, e.g. VLT, XMM,
CHANDRA and PLANCK, will certainly make cosmology with clusters 
an even more fascinating field in the future.

\acknowledgements
I acknowledge gratefully the hospitality of the Institut d'Estudis
Espacials de Catalunya in Barcelona where these proceedings were
written. During 
the stay there I was supported by the TMR grant ERB-FMGE CT95 0062 by
CESCA-CEPBA.

\end{document}